# An erbium-doped waveguide amplifier on thin film lithium niobate with an output power exceeding 100 mW

Rui Bao[1,2], Zhiwei Fang [1,*], Jian Liu[1], Zhaoxiang Liu[1], Jinming Chen[1], Min Wang[1], Rongbo Wu[1,*], Haisu Zhang[1,*], and Ya Cheng[1,2,3,4,5,6,7,*]

[1] The Extreme Optoelectromechanics Laboratory (XXL), School of Physics and Electronic Science, East China Normal University, Shanghai 200241, China

[2] State Key Laboratory of Precision Spectroscopy, East China Normal University, Shanghai 200062, China

[3] State Key Laboratory of High Field Laser Physics and CAS Center for Excellence in Ultra-intense Laser Science, Shanghai Institute of Optics and Fine Mechanics (SIOM), Chinese Academy of Sciences (CAS), Shanghai 201800, China

[4] Collaborative Innovation Center of Extreme Optics, Shanxi University, Taiyuan 030006, China

[5] Collaborative Innovation Center of Light Manipulations and Applications, Shandong Normal University, Jinan 250358, People's Republic of China

[6] Hefei National Laboratory, Hefei 230088, China

[7] Joint Research Center of Light Manipulation Science and Photonic Integrated Chip of East China Normal University and Shandong Normal University, East China Normal University, Shanghai 200241, China

* zwfang@phy.ecnu.edu.cn; rbwu@phy.ecnu.edu.cn; hszhang@phy.ecnu.edu.cn; ya.cheng@siom.ac.cn

**We demonstrate high-power thin film lithium niobate (TFLN) erbium-doped waveguide amplifier (EDWA) with a maximum on-chip output power of 113 mW and a gain of 16 dB. The on-chip integrated EDWA is composed of large mode area (LMA) waveguide structures with a total length of 7 cm and a footprint of 1×1 cm$^2$. Particularly, we connect segmented LMA waveguides with waveguide tapers to achieve on-chip mode conversion which maintains single-mode propagation all over the EDWA even at the waveguide bends. The design leads to significant increase of the amplified signal power by orders of magnitude and will open an avenue for applications such as on-chip high-power lasers and amplifiers system.**

## 1. Introduction

As the backbone of the global telecommunication network, erbium doped fiber amplifier (EDFA) plays an indispensable role in the high-capacity intercontinental data transmission since its first invention in late 1980s [1]. The stable optical transitions of erbium ions featuring a long-excited state lifetime give rise to a broadband gain covering the conventional telecom band (C-band). The unique characteristic makes EDFA ideally suitable for high-bit-rate multiwavelength amplification. The great success of erbium doped fibers boosts continuous efforts on embedding erbium ions into other waveguide platforms for miniaturized amplifiers, i.e., erbium doped waveguide amplifiers (EDWA) [2]. An abundance of host materials and waveguide configurations have been investigated in the past decades, including the ion-exchanged/diffused bulk waveguide in silicate glass and lithium niobate crystal, lithographically patterned planar waveguide in optical thin films like $Al_2O_3$, $Si_3N_4$ and $Ta_2O_5$ [3]-[7]. Nevertheless, to realize a compact EDWA of high scalability, high gain and high output power, stringent requirements on the waveguide propagation loss, erbium doping concentration, optical confinement and modal overlap must be fulfilled at once. So far, EDWA of output powers above 100 mW has only been realized in ultralow-loss erbium implanted $Si_3N_4$ waveguide amplifiers [7].

The advent of high-quality thin film lithium niobate (TFLN) on insulator and the advancement of microfabrication technique promotes the rapid development of TFLN photonic integrated circuits, benefited from the excellent nonlinear optical and electro/acousto-optical response of lithium niobate crystal as well as the tight optical confinement in TFLN waveguide facilitating dense integration and high interaction strength [8]-[10]. Recent demonstration of erbium-doped TFLN (Er: TFLN) wafer and integrated waveguide amplifiers and micro-lasers puts forward another paradigm for efficient EDWA applications, since the TFLN waveguides nominally support high-weight erbium doping due to the adequate solvability of the LN matrix, tight optical confinement by the large refractive index contrast and ultralow propagation loss (as low as 1 dB/m) when fabricated by the suitable photolithographic technique [11]-[20]. Nevertheless, the established TFLN EWFAs until now are rather limited in the attainable gain and output powers, i.e., the output powers of previously recorded Er: TFLN waveguide amplifiers are saturated at around 1 mW with the power conversion efficiency below 1%. To overcome this issue, it is known that in conventional EDFAs, large mode area (LMA) fibers are commonly used to achieve high-power output [21]-[26].

In this work, by use of a LMA waveguide structure, we overcome the gain and power deficiency of TFLN-based EDWA and achieve an on-chip output power above 100 mW and a power conversion efficiency greater than 50%. The LMA design features an array of wide multimode waveguides connected to nearly single-mode waveguides by adiabatic tapers for fundamental mode excitation as well as compact waveguide bends which can maintain the single-mode propagation in successive LMA waveguides. The LMA waveguide amplifier, which holds a cross-section of $0.8\times9$ $\mu m^2$ and extends 7 cm in a compact footprint of $1\times1 cm^2$, is fabricated using our home-developed photolithography assisted chemo-mechanical etching (PLACE) technique. Experimental characterization of the Er: TFLN waveguides with different widths reveals the advantage of LMA in promoting amplifier output power and conversion efficiency, which is theoretically reproduced by the amplifier model including the steady-state response of erbium ions subject to excited state absorption. Simulation results pinpoint the population trapping of erbium ions in LN crystals as the potential intensity dependent nonlinear impairment which can be greatly relaxed through the LMA waveguide design. The noise figures and the parasitic lasing effect of the fabricated amplifier chip are also characterized, both of which can be mitigated using advanced edge couplers to realize a low noise self-lasing-free high-power EDWA.

## 2. Results and discussion

In this work, the on-chip EDWA is fabricated on an 800-nm-thick Z-cut $Er^{3+}$-doped thin film lithium niobate (Er: TFLN) substrate. The design of LMA waveguide amplifier is shown in Fig. 1a, in which the waveguide configuration is featured with a series of narrow waveguides and wide waveguides interconnected using adiabatic tapers. The input and output ports, as well as the bending regions, all utilize narrow waveguides with a width of 1 μm. The main gain region, on the other hand, employs LMA waveguides with a width of 9 μm. The input and output ports with 1-μm-wide waveguide are primarily used for efficient light coupling using the lensed fibers. The 1-μm-wide bending waveguides are primarily used to filter out undesirable higher-order modes and obtain a compact footprint. The adiabatic waveguide tapers are used for transforming the tight mode to the large mode. The three figures on the right-handed side of Fig. 1a respectively display the simulated transverse-electric (TE) mode field profiles of three different waveguide widths: 1 μm, 3 μm, and 9 μm. A simplified $Er^{3+}$ energy level diagram in Fig.1a shows 1550-nm signal wavelength is amplified when a 1480-nm pump wavelength is used. Fig. 1(b) shows the numerical calculation of the field intensity in the tapers at a wavelength of 1550 nm. The input light source was set as the fundamental TE-mode. The adiabatic modal evolution in the waveguide taper can be clearly observed in Fig. 1(b), which is crucial for single mode propagation within the EDWA chip. Fig. 1(c) shows the optical microscope image of the fabricated TFLN waveguide taper with a transition length of the 1 mm. We fabricate a 7-cm-long EDWA on TFLN using our home-developed PLACE technique [27]-[29]. Fig.1(d) shows the digital camera image of the fabricated Er: TFLN waveguide amplifier chip with a compact footprint of 1×1 $cm^2$ butt-coupled with two lensed fibers. Bright green fluorescence from erbium ions under 1480-nm optical pumping is clearly observed in the EDWA chip. Microscopic top-view images of three waveguide sections with different top-widths are shown in Fig.1 (e-g), where excellent fundamental mode propagation even in the LMA waveguides can be convincingly observed.

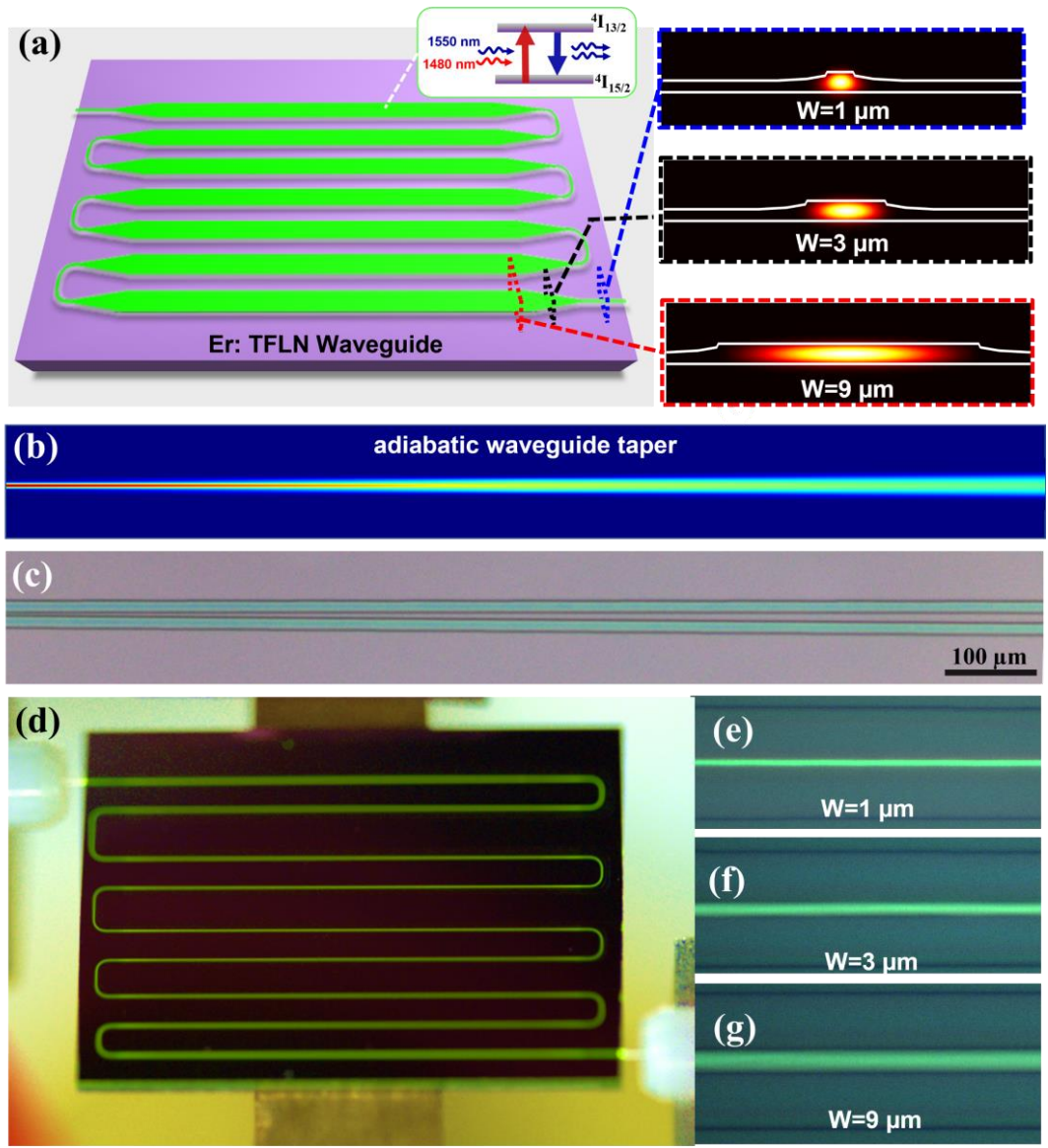

**Figure 1. On-chip Er: TFLN waveguide amplifier.** (a) Schematic of the on-chip Er: TFLN waveguide amplifier, the insets show a simplified energy diagram of erbium-doped LN with the pump at 1480 nm and the signal at 1550 nm. The mode images around the input/output ports and the waveguide bends, in the transition region of the waveguide taper and the LMA waveguide are shown in the insets. (b) Simulated field intensity for adiabatic waveguide tapers at a wavelength of 1550 nm. (c) Optical image of a fabricated 1-mm-long adiabatic waveguide taper. (d) Optical image of a 7-cm-long Er: TFLN waveguide amplifier chip butt-coupled with two lensed fibers. Green up-conversion fluorescence was observed in waveguides of different widths: (e) 1 μm, (f) 3 μm and (g) 9 μm.

First, the critical impact of the waveguide mode-area on the amplifier performance is characterized. A series of straight waveguides of the length of 6 cm and varying top-widths (1 μm~15 μm) are fabricated for experimental examination. All the waveguides have the same top-width of 1 μm at the input and output ends using adiabatic waveguide tapers for uniform coupling with the lensed fibers. The coupling loss between the lensed fiber and the waveguide facet is measured to be 6.5 dB at the wavelength of 1550 nm. The amplifier gain and output power of the fabricated erbium-doped waveguides are characterized by the pump-probe method using the experimental setup shown in Fig. 2(a). Two 1480 nm high-power fiber-coupled diode lasers are employed as the pump source, and a 1550 nm tunable external cavity diode laser (Toptica) is used as the signal source. The maximum output power of each pump diode laser is 500 mW. The pump laser and signal laser are first combined by the fiber-based wavelength division multiplexer (WDM) and then co-injected into the waveguide chip through the lensed fiber with the cleaving angle of 99°. A second fiber-WDM is used at the output end of the waveguide chip for injecting the other pump laser and extracting the amplified signal at the same time. Several in-line polarization controllers (PCs) are inserted in the fiber pass of the pump and signal lasers for input polarization adjustment. Fundamental TE-modes are excited in the waveguide chip for reducing propagation loss and enhancing erbium absorption and emission rates compared to the transverse-magnetic (TM) modes in our Z-cut Er: TFLN wafer.

The off-chip output signal powers are measured by the optical spectrum analyzer (OSA) and then converted to the on-chip output power assuming the coupling loss of 6.5 dB per facet. The on-chip pump power is recorded in the same way by measuring the input power before the waveguide facet and then deducting the facet coupling loss. The on-chip signal output powers of the waveguides with 1 μm and 9 μm top-width at increasing on-chip pump powers are plotted in Fig. 2 (b) as orange squares and blue triangles, respectively. The input signal power is fixed at 12 dBm, corresponding to 3.5 mW on-chip input signal power. It can be clearly seen from Fig. 2(b) that for the 1-μm-wide waveguide, the output signal power is rapidly saturated at 18 mW with the on-chip pump power of 55 mW and then the output power gradually drops when the on-chip pump power is increased to above 200 mW. In contrast, for the 9-μm-wide waveguide, the output signal power increases quasi-linearly with the on-chip pump powers until reaching the maximum on-chip pump power of 210 mW, at which a maximum on-chip output power of 113 mW is obtained corresponding to the power conversion efficiency (PCE) of 52%. Such sharp difference between the amplifier output power of the narrow and wide waveguides pinpoints the significant role of waveguide mode areas, implying the intensity-dependent nonlinear impairment as the limiting effect in the Er: TFLN waveguide amplifiers.

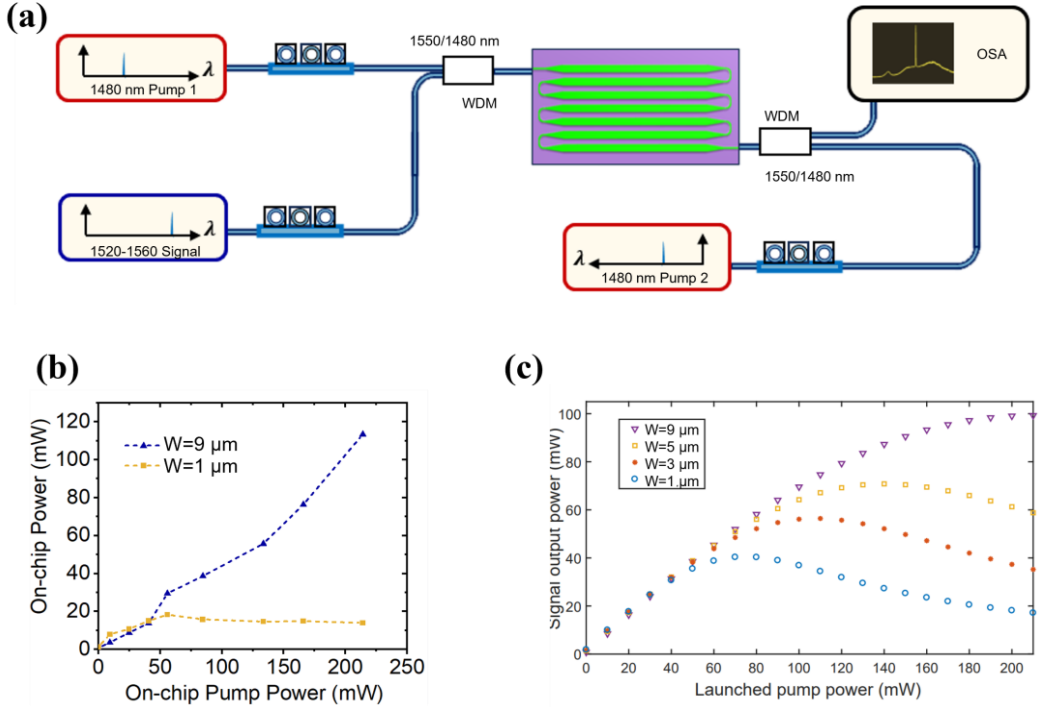

**Figure 2. Experimental characterization**. (a)Experimental schematic for the waveguide amplifier characterization. (b)The measured on-chip signal power with respect to the on-chip pump power. (c)The simulated amplifier output power vs the launched pump power.

To gain further insight on the observed waveguide amplifier performance, the steady-state amplifier model considering the population dynamics of erbium ions and the modal field distribution of the waveguide is adopted to simulate the amplifier with different waveguide mode-areas. Three energy levels of erbium ions $^4I_{15/2}$, $^4I_{13/2}$, $^4I_{11/2}$ are explicitly included, and the absorption and stimulated emission between $^4I_{15/2}$ and $^4I_{13/2}$ by the pump laser and signal laser are described by the wavelength-dependent cross sections of erbium ions characterized in the bulk-doped lithium niobate crystals. Cooperative up-conversion (CUC) between excited erbium ions in the level of $^4I_{13/2}$ and $^4I_{11/2}$ are also included. Moreover, due to the high laser intensity in the tightly-confined TFLN waveguide (10 mW in the 1-μm-wide waveguide corresponds to the peak intensity of $10^6$ W/cm$^2$), excited-state-absorption (ESA) from the $^4I_{13/2}$ level is considered which can play an adverse effect by excited-state depletion at very high pumping intensities [30].

The obtained simulation results are depicted in Fig. 2(c), where the same input signal and pump powers with experiment are chosen in the simulation. The output signal powers feature identical behaviors as the saturated output power is reduced with the smaller waveguide mode-areas and the remarkable descending output powers at high pump powers for the narrower waveguide. The simulated results qualitatively reproduce the experimental observation, which corroborate the suitability of the employed amplifier model. Detailed analysis reveals the significant contribution of excited-state absorption from $^4I_{13/2}$ to $^4I_{9/2}$ at high pump intensities, with the latter $^4I_{9/2}$ level rapidly decaying by nonradiative multiphoton relaxation to $^4I_{11/2}$ level whose lifetime is relatively long (~200 μs) [31]. Such process depletes the population inversion between $^4I_{13/2}$ and $^4I_{15/2}$ by population trapping in $^4I_{11/2}$, causing the reduced erbium gain and output signal power at high pump powers in narrow waveguide [32][33]. The revealed physics evidences the superiority of wide waveguide with LMA for high-efficiency amplification and power conversion.

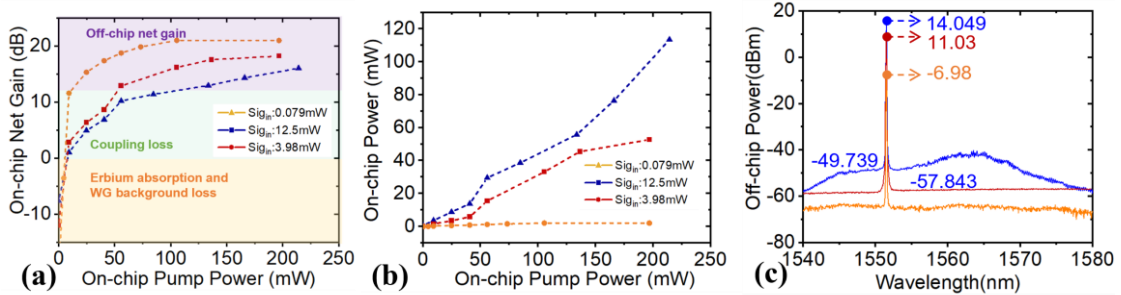

**Figure 3. Optical characterization of the amplifier chip**. (a)The on-chip net gain of the fabricated amplifier chip with respect to the on-chip pump power at three input signal powers. (b)The on-chip output signal power of the fabricated amplifier chip with respect to the on-chip pump power at three input signal powers. (c)The input and output spectra of the amplifier chip showing the background spectral noise.

After performing the systematic investigation on the mode-area effect of waveguide amplifier, a compact waveguide amplifier chip based on Z-cut Er: TFLN wafer is fabricated with double-folded footprint for dense integration (like the one shown in Fig. 1(a)). The straight segments of the amplifier chip are all 9-μm-wide waveguides, while the curved segments are Euler-shaped waveguide bends of 1 μm top-width. Adiabatic waveguide tapers are employed for the transition part between the 9-µm-wide and 1-µm-wide waveguides. Gain performance of the amplifier chip is first characterized. To avoid the overestimate of the on-chip gain by the signal enhancement method, the off-chip signal input and output powers are directly measured and calibrated. Then the on-chip signal input power and output power are deduced from the off-chip power by reasonably assuming the same coupling loss at both ends of the amplifier chip.

The measured amplifier gain response at three different signal input powers is shown in Fig. 3(a). The input signal wavelength is set at 1550 nm. The maximum small signal on-chip and off-chip gains are 20.5 dB and 7.5 dB, respectively. Besides, the measured output signal powers on the amplifier chip are also shown in Fig. 3(b), where a maximum power of 110 mW is achieved at the on-chip pump power of 210 mW, in consistent with the results shown in Fig. 2(b) obtained from a long and straight 9-µm-wide waveguide. It thus indicates that multiple waveguide bends in the compact amplifier chip play little effect on the gain process, corroborating the advantage of large mode-area amplification in the wide waveguide rather than the narrow waveguide.

The output noise figure of the amplifier chip is further characterized by the optical source subtraction method. The source spontaneous emission (SE) noise of the input signal laser is measured and removed from the total noise arising from the amplifier chip. The noise figure of the amplifier chip can be deduced by the equation shown below [7]:

$$\text{NF} = \frac{P_{\text{out}} - GP_{\text{SE}}}{Gh\nu B_0} + \frac{1}{G}$$

where $P_{\text{out}}$ is the output signal power, $P_{\text{SE}}$ is the input signal spontaneous emission noise power, $G$ is the amplifier gain, $h$ is the Planck constant, $\nu$ is the signal frequency and $B_0$ is the optical integration bandwidth (the resolution of 0.05 nm is employed in the OSA spectrum measurement).

As shown in Fig. 3(c), the measured input signal spectrum is plotted in red while the amplified output signal spectrum is depicted in blue. The output signal spectrum without pump is also shown in orange color for comparison. The input signal power is around 11 dBm, while the output signal power is 14 dBm. Both values are measured in the input and output fibers. At the 3 dB off-chip gain (16 dB on-chip gain), the input SE-noise at the signal wavelength of 1552 nm is interpolated to be -57.8 dBm and the output amplifier noise at the signal wavelength is fitted to be -49.7 dBm. Then the noise figure of the amplifier chip is calculated to be 7.1 dB, which is large compared to the conventional EDFAs due to the excess fiber-chip coupling losses. The noise figure of erbium-doped TFLN amplifier can be further suppressed using advanced coupling interface such as edge couplers and spot size convertors.

The amplified spontaneous emission (ASE) spectrum of the amplifier chip is further shown in Fig. 4(a). A broad emission band of erbium ions ranging from 1510 nm to 1590 nm with the highest peak around 1531.5 nm facilitates the wideband amplification in the fabricated amplifier chip. In the high-gain regime, the background ASE noise will be reinforced by the back-reflections at the input and output waveguide facets, and parasitic lasing at the peak fluorescence wavelength of 1531.5 nm sets in when the on-chip round trip gain compensates the round-trip loss determined by the facet reflection and background propagation loss. The small-signal gain measurements at the input signal wavelengths of 1531.5 nm and 1550 nm are shown in Figs. 4(b) and 4(c), respectively. At the 1531.5 nm input signal wavelength, the parasitic lasing is clearly observed when the amplifier on-chip gain is higher than 28 dB, while the parasitic lasing emerges earlier for the 1550 nm input signal wavelength, i.e., parasitic lasing is remarkable when the amplifier on-chip gain is higher than 21 dB. Moreover, the small signal amplification at the 1550 nm is largely disturbed in the parasitic lasing regime, due to the smaller emission cross section of erbium ions at the signal wavelength compared to that at the lasing wavelength. The parasitic lasing by facet back-reflections can be further optimized using angled waveguide facet and index matched edge couplers, which will be systematically investigated in the future.

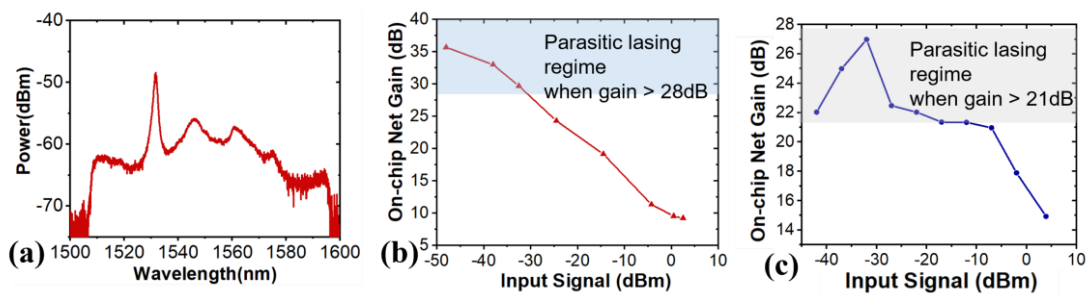

**Figure 4. Gain saturation and parasitic lasing**. (a)The amplified spontaneous emission spectrum of the amplifier chip. (b)The on-chip net gain of the amplifier chip with respect to the input signal power at the wavelength of 1531.5 nm. (c)The on-chip net gain of the amplifier chip with respect to the input signal power at the wavelength of 1550 nm.

## 3. Conclusion

To conclude, a high-power compact EDWA is demonstrated in the erbium-doped TFLN platform employing the LMA waveguides interconnected with adiabatic waveguide tapers and bends. The on-chip output powers of the fabricated EDWA chip are measured to be higher than 100 mW at the C-band, with the corresponding on-chip net gain of 16 dB and power conversion efficiency of 52%. In a nutshell, the results are achieved with the combined contributions of large mode area, low propagation loss, nearly single-mode propagation, and high doping concentration in the EDWA device which can be mass-produced on TFLN using the PLACE technique.